%% file: template-A4.tex
\documentclass[conference]{IEEEtran}
\IEEEoverridecommandlockouts
\usepackage{cite}
\usepackage{amsmath,amssymb,amsfonts}
\usepackage{algorithm,algpseudocode}
\usepackage{graphicx}
\usepackage{textcomp}
\usepackage{booktabs}
\usepackage{xcolor}
\usepackage[normalem]{ulem}

\usepackage[a4paper, total={184mm,239mm}]{geometry}
\def\BibTeX{{\rm B\kern-.05em{\sc i\kern-.025em b}\kern-.08em
    T\kern-.1667em\lower.7ex\hbox{E}\kern-.125emX}}
\usepackage{subcaption}
\usepackage{array,mathtools, multirow}
\input{math_commands.tex}

\usepackage{color}

\usepackage[capitalise]{cleveref}
\begin{document}

\title{RADAR: Run-time Adversarial Weight Attack Detection and Accuracy Recovery

}
\author{\IEEEauthorblockN{Jingtao Li\IEEEauthorrefmark{1}, Adnan Siraj Rakin\IEEEauthorrefmark{1}, Zhezhi He\IEEEauthorrefmark{2}, Deliang Fan\IEEEauthorrefmark{1}, Chaitali Chakrabarti\IEEEauthorrefmark{1}}
\IEEEauthorblockA{\IEEEauthorrefmark{1}School of Electrical Computer and Energy Engineering,
Arizona State University, Tempe, AZ, 85287}
\IEEEauthorblockA{\IEEEauthorrefmark{2}Department of Computer Science and Engineering, 
Shanghai Jiao Tong University, Shanghai}
\IEEEauthorrefmark{1}\{jingtao1, asrakin, dfan, chaitali\}@asu.edu; \IEEEauthorrefmark{2}\{zhezhi.he\}@sjtu.edu.cn
}
\maketitle

\begin{abstract}
Adversarial attacks on Neural Network weights, such as the progressive bit-flip attack (PBFA), can cause a catastrophic degradation in accuracy by flipping a very small number of bits. Furthermore, PBFA can be conducted at run time on the weights stored in DRAM main memory. In this work, we propose RADAR, a \underline{R}un-time adversarial weight \underline{A}ttack \underline{D}etection and \underline{A}ccuracy \underline{R}ecovery scheme to protect DNN weights against PBFA. We organize  weights that are interspersed in a layer into groups and employ a checksum-based algorithm on weights to derive a 2-bit signature for each group. At run time, the 2-bit signature  is computed and compared with the securely stored golden signature to detect the bit-flip attacks in a group. After successful detection, we zero out all the weights in a group to mitigate the accuracy drop caused by malicious bit-flips. The proposed scheme is embedded in the inference computation stage. For the ResNet-18 ImageNet model,  our method can detect 9.6 bit-flips out of 10 on average. For this model, the proposed accuracy recovery scheme can restore the accuracy from below 1\% caused by 10 bit flips to above 69\%. The proposed method has extremely low time and storage overhead. System-level simulation on gem5 shows that RADAR only adds $<$1\% to the inference time, making this scheme highly suitable for run-time attack detection and mitigation. 

\end{abstract}


\section{Introduction}
Neural networks have been widely adopted in image recognition, natural language processing, medical diagnosis and autonomous driving tasks. The security and trustworthiness of neural networks directly affect the safety of these applications making this study  even more important.
Neural network models have been shown to be vulnerable to various types of attacks. 
Adversarial input attack, which manipulates the inputs fed to the neural network model, such as FGSM \cite{goodfellow2014explaining}, can cause serious misclassification. Recently, adversarial weight attack with malicious weight bit-flips, aka. PBFA \cite{rakin2019bit}, on ResNet-18 model was able to  degrade ImageNet classification accuracy to below 0.2\% with only 13 bit-flips. 
Furthermore, \cite{yao2020deephammer} showed how weight attacks can be mounted at run-time to circumvent protection schemes that perform detection  periodically.

There is only a handful of techniques that provide some level of  security against adversarial weight attacks. The passive defense method in \cite{hedefending2020} applies regularization to make the weights more resistant to weight attacks. However it is incapable of detecting whether an attack has occurred or not.
Error correction code (ECC) based schemes proposed in \cite{qin2017robustness,guan2019place} provide protection against random soft-errors but not against  adversarial attacks. Standard data integrity checking methods such as MD5, CRC can perform detection but with high overhead. 
These are generic techniques and do not exploit the characteristics of the neural network model or the specifics of the attacks that are launched against the networks. 

As a countermeasure to PBFA, we propose RADAR, a Run-time adversarial weight Attack Detection and Accuracy Recovery scheme. It operates on weights that are fetched from DRAM to on-chip cache for inference computation. 
RADAR leverages the PBFA characteristics to derive a simple checksum based technique that has excellent error detection and accuracy recovery performance.
The weights in a layer are reorganized into groups for the checksum computation, where each group has weights that were originally $k$ locations apart, $k>1$.  The checksum is computed on the weights in  a group that have been masked using a secret key and is used to derive a 2-bit signature. At run-time, the 2-bit signature of a group is compared with the secure signature to detect possible errors. 
Once an error is flagged, all weights in that group are replaced with zeroes. The storage and the time overhead of this method is very small compared to the RADAR-free inference baseline.
Our contributions can be summarized as follows:
\begin{itemize}
\item We present RADAR, a low latency and storage overhead run-time scheme that can provide effective detection and recovery on state-of-the-art adversarial weight attack on DNN, namely, Progressive Bit-Flip Attack (PBFA).
\item RADAR computes addition checksum on a group of masked weights that were originally interspersed  to derive a 2-bit signature. Use of interleaved weights and masking  helps achieve  a high detection ratio of 96.1\% on a 10-bit PBFA attack on ResNet-18 model. 
\item RADAR employs a simple scheme where all weights in a group are set to zero if that group has been flagged with an error. For ResNet-18 on ImageNet, the accuracy drop due to PBFA can be recovered from 0.18\% to greater than 60\% when the partition size is 512.
\item System-level Gem5 simulations show that the time cost of RADAR is $<$1\% of the inference time for ResNet-18.
The overhead to store the signature is only 5.6 KB for ResNet-18, making it feasible to be stored securely on-chip.
\end{itemize}

\vspace*{-0.05in}
\section{PRELIMINARIES}




\subsection{DNN Attack, Defense \& Detection.}

Recent developments of  memory fault injection techniques on hardware~\cite{kim2014flipping, agoyan2010flip} have made directly attacking model parameters, such as DNN weights, at run time feasible.
Among them, row-hammer attack which causes bit-flips in Dynamic Random-Access Memory (DRAM) through repeatedly activating DRAM rows, is the most popular one~\cite{kim2014flipping, gruss2018another,255272}.
Adversarial weight attack~\cite{hong2019terminal,liu2017fault,rakin2019bit} corrupts the neural network weights directly to achieve certain attacking goals~\cite{liu2017fault,rakin2019bit}. 
A recently developed adversarial weight attack, known as Progressive Bit-Flip Attack (PBFA), identifies vulnerable bits based on gradient information and degrades DNN classification accuracy to random guess level~\cite{rakin2019bit}.

To mitigate affect of adversarial weight attacks, attack defense mechanisms have also been investigated~\cite{hedefending2020}. For instance, \cite{hedefending2020} uses binarization or a relaxed version of the binarization technique to handle PBFA attacks. This method increases the resistance to the PBFA attack and is a good passive defense method.  
DNN soft error detection schemes can be used to detect small perturbations in weights~\cite{he2019sensitive}. 
Error Correction Codes (ECC)-based techniques~\cite{qin2017robustness, guan2019place,8388830}  have been shown to correct soft errors in neural network models. However, rowhammer attack can be used to compromise ECC codewords in DRAM, making these methods not as effective.

\section{Threat Model}
\begin{figure}[ht]
  \centering
  \includegraphics[width=\linewidth]{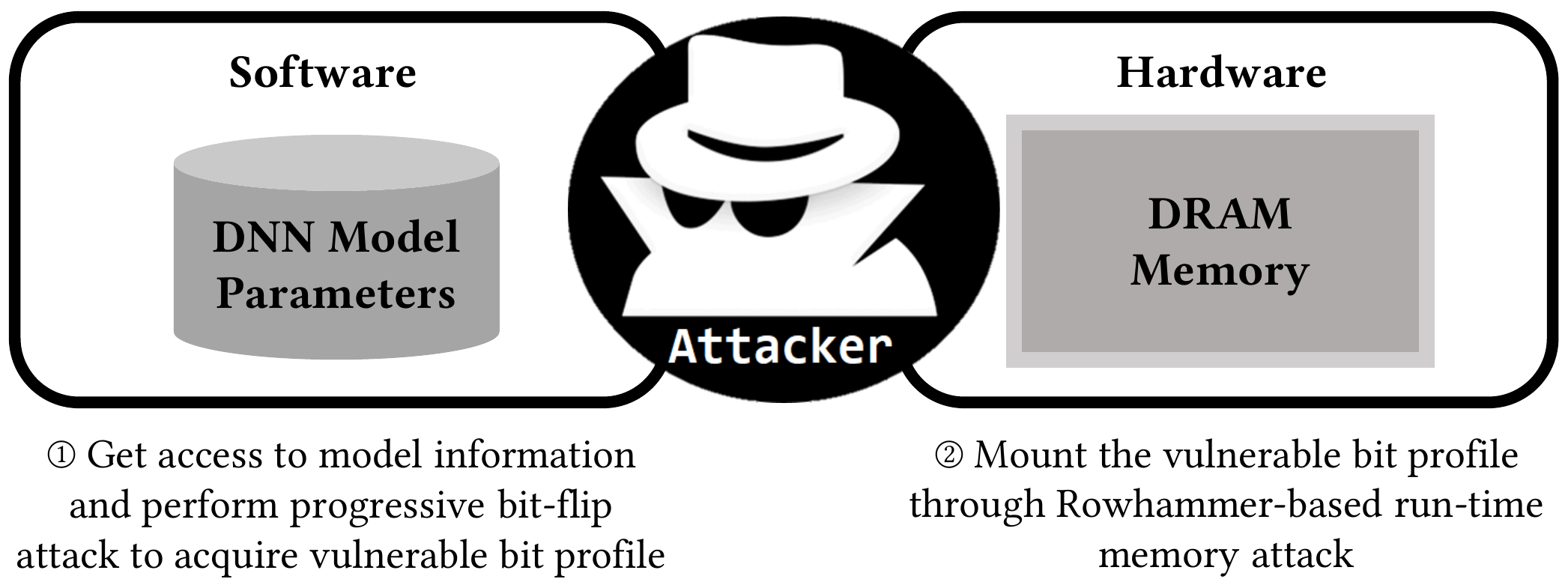}
  \caption{Software and hardware aspects of the threat model.}
  \label{fig:fig1}
\end{figure}

\cref{fig:fig1} describes the threat model in this work.
At the software end, the attacker uses PBFA in~\cite{rakin2019bit} to identify the vulnerable bits, and at the hardware end, the attacker performs fault injection via DRAM row-hammer attack by mounting the vulnerable bits at run-time, thus corrupting the stored weights. We consider DNNs with 8-bit quantized weights as in~\cite{rakin2019bit}.

\subsection{Hardware Assumptions}
Row-hammer attack has been demonstrated to be very effective in corrupting  DRAM contents~\cite{kim2014flipping, gruss2018another, 255272}. The neural network weight parameters are are very large in size (MB$\sim$GB) and hence stored in DRAM. 
Recent work in~\cite{255272} has demonstrated how DRAM weights can be attacked using rowhammer  in practice. 



We consider all weight attacks are physically implemented by DRAM row-hammer attack on PBFA identified vulnerable weight bits. 
Since every bit flip attack costs time and effort, we assume that the attacker stops the bit flip attacks after causing a significant accuracy drop.
We also assume that the attacker is unable to attack data stored securely in on-chip SRAM.  
Additionally we assume that the attacker cannot corrupt the system kernels (otherwise the attacker would be able to break the system~\cite{seaborn2015exploiting}). 


\subsection{Software Assumptions}

We only consider Progressive Bit-Flip Attack (PBFA)~\cite{rakin2019bit} since it is the strongest adversarial weight attack technique to date. It causes the DNN to malfunction with the fewest number of bit-flips. We argue performing random bit-flip is too weak to be considered as an attack. It has already been demonstrated in~\cite{rakin2019bit} that randomly flipping 100 bits merely degrades the accuracy by less
than 1\%.

To perform PBFA, the attacker has access to the network architecture and parameters, e.g., weight, bias, etc. (white box assumption). Such information can be acquired by acting as a benign user or revealed through side-channels \cite{yudeepem}. 
To perform BFA, we assume the attacker has a small dataset with roughly similar distribution as the training data to get accurate gradient information. Additionally, we assume that the attacker has some knowledge of the defense mechanism (aka checksum) but does not know of the secret key used for generating masked weights or the interleaving strategy.
\subsection{Characteristics of PBFA.} 
PBFA is a very powerful attack that can severely degrade the accuracy with only a few bit flips.
Our experiments show that, on average, with 10 bit-flips, the accuracy of a trained 8-bit ResNet-20 model on CIFAR-10 dataset can drop from above 90\% to 18.01\% and the accuracy of an 8-bit ResNet-18 model on ImageNet can drop from around 70\% to 0.18\%.

 
To derive an effective detection scheme for PBFA-based attacks, we first did an in-depth characterization of the attack. We generated multiple sets of PBFA bit profiles and did a statistical analysis.
Specifically, we performed 100 rounds of PBFA with 10 bit-flips per round on ResNet-20 model and ResNet-18 model, saved the profiles of the vulnerable bits in each round, and computed the statistics. 
\begin{table}[ht]
\centering
\vspace{-5pt}
  \caption{Number of PBFA Attacks in Different Bit Positions over 100 rounds}
  \label{tab:BFA_pattern}
  \begin{tabular}{|c|c|c|c|}
    \hline
    {}  & MSB (0 $\rightarrow$ 1) & MSB (1 $\rightarrow$ 0) & others\\
    \hline
    ResNet-20 &334 &666 &0\\
    ResNet-18 &16  &897 &87\\
    \hline
  \end{tabular}
  \vspace{-5pt}
\end{table}

\textbf{Observation 1.} The PBFA attack exploits the non-uniformity in the importance of some bits over others in a quantized representation. PBFA always chooses to flip the Most Significant Bit (MSB) in a weight.  \cref{tab:BFA_pattern} shows that  the MSBs are targeted (334+666)/1000 times for ResNet-20 and (16+897)/1000 times for ResNet-18. Thus a low overhead detection scheme should target detecting bit-flips in the MSB position.

\begin{figure}[t]
  \centering
  \includegraphics[width=1.0\linewidth]{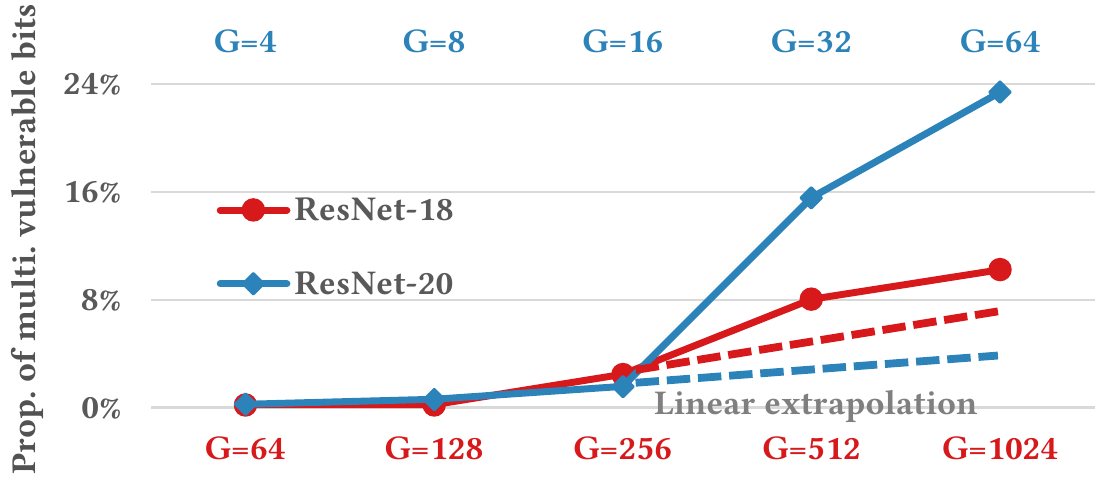}
  \vspace{-1em}
  \caption{Proportion of occurrences of multiple vulnerable bits in the same group.}
  \label{fig:figA}
  \vspace{-5pt}
\end{figure}

\textbf{Observation 2.} The vulnerable bits identified by PBFA have a scattered spatial distribution. In this experiment, we partition the weights into groups with $G$ weights in a group, and count the number of bits picked by PBFA in each group.
Fig. \ref{fig:figA} shows that the proportion of multiple vulnerable bits inside one group is very low when $G$ is small (relative to the model size), and the proportion grow in a super-linear manner for larger group sizes. This indicates that vulnerable bits are scattered across groups rather than being clustered in a group.


\textbf{Observation 3.} The bit-flip locations in PBFA are more likely to occur on weights that have very small values. 
As shown in Table~\ref{tab:BFA_weight1}, most of the bit-flips happen on weights that have values in the range (-32, 32). Thus, after the bit-flip, the weight value will be in either (96, 127) or (-128, -96) range. We believe that the large weight change is the main cause of severe accuracy drop in PBFA attacks.



\begin{table}[ht]
\vspace{-5pt}
  \caption{Frequency of targeted weights in different ranges}
  \label{tab:BFA_weight1}
  \centering
  \begin{tabular}{|c|c|c|c|c|}
    \hline
    Range & (-128, -32)& (-32, 0) &(0, 32)& (32, 127)\\
    \hline
    ResNet-20  &85 &595 &249 &71\\
    ResNet-18 &16 &860 &76 &27\\
    \hline
  \end{tabular}
  \vspace{-5pt}
\end{table}

\section{RADAR Scheme: Detection}


We assume that the weight values are loaded from DRAM main memory into on-chip caches and then 
processed. 
A well-designed computing scheme maximizes the weight reuse so that each weight is accessed only once. 
Since the main memory is vulnerable to rowhammer attack, the weights stored there could be compromised and so detection has to be performed on all weights that are loaded into cache prior to processing.  


In order to embed detection in the inference process, the proposed method has to have the following properties:
\begin{itemize}
    \item Low timing overhead. The time to perform detection adds to the total inference computation time and thus has to be as small as possible. 
    \item Low storage overhead. The golden signature that is required for detection has to be small enough to be stored in the secure on-chip memory.
\end{itemize}



\subsection{Checksum-based Signature Calculation}
Popular detection schemes based on CRC or SEC-DED have high storage overhead (Section~VII.B) and are not applicable. We adopt an addition-based checksum scheme~\cite{maxino2009effectiveness} for its simplicity, and add interleaving of weights and checksum on masked weights to improve the attack resistance. Specifically, we
compute $M$, the sum of $G$ weights in a group and derive a two-bit signature $S_{i,j} = \{S_A, S_B\}$ from $M$ for $i$-th layer, $j$-th group in the following way:
\begin{equation}
S_A = \lfloor M/256 \rfloor \% 2; \hspace*{0.5in}
S_B = \lfloor M/128 \rfloor \% 2
\label{eq:eq1}
\end{equation}





In equation \ref{eq:eq1}, the $\lfloor \cdot \rfloor$ denotes the floor function and $\%$ denotes the remainder function. Note that the binarization step can be simply implemented as bit truncation in hardware.
Similar to the parity code, $S_{B}$ can detect any odd number of bit-flips 
on MSBs of a group of $G$ weights. From Fig.~\ref{fig:figA} we see that most groups have single bit-flips which can be detected by the parity bit $S_{B}$ 100\% of the time. Also, when the group size is large, multiple bits in a group could be flipped.
Since $S_B$ is blind to any even number of bit-flips, we include a second bit, $S_A$, 
which can only detect double bit-flips if they occur in the same direction, i.e., the bit-flips are of the form
(0$\rightarrow$1, 0$\rightarrow$1) or (1$\rightarrow$0, 1$\rightarrow$0). However, flips of the form (0$\rightarrow$1, 1$\rightarrow$0) cannot be detected since they do not change the value of $M$. Next we show how this weakness can be addressed by computing the checksum on weights that are masked and interleaved. We argue that it is less efficient to incorporate more bits into the signature, such as one more bit to protect the MSB-1 position by computing $S_C=\lfloor M/64 \rfloor$. This is because attacking MSB-1 would require the attacker to flip more bits to achieve the same attacking performance as discussed in section VIII.




\subsection{Improving attack detection capability of checksum} 
We adopt the following two features to improve the attack detection capability of the simple addition checksum approach.\\
\textbf{1. Masking Weights in Checksum Computation:}  We argue that simply performing addition checksum to derive the signature makes the system risky. We use a randomly generated secret key as a mask on a group of weights to determine whether or not to take its two's complement or not during the summation (lines 4-6 of Algorithm 1). 


The secret key is $N_k$ bits long and changes from layer to layer. Increasing $N_k$ can reduce the probability of the sequence of operations being guessed correctly by the attacker but comes with a high implementation cost. We  set $N_k = 16$, and the $2^{16}$ different combinations  provide for sufficient security. \\
\textbf{2. Interleaving Weights for Checksum Computation.} 
Given the double bit error blindness of addition checksum, the attacker can intentionally target multiple bits in the same group to attack in order to bypass the detection.
So we compute the checksum on a group of weights, where the weights in a group were originally $m$ locations apart, $m>1$. This process is referred to as interleaving, a well known technique that is used to handle detection of burst errors in communication systems.

\begin{figure}[ht]
  \vspace{-5pt}
  \centering
  \includegraphics[width=0.7\linewidth]{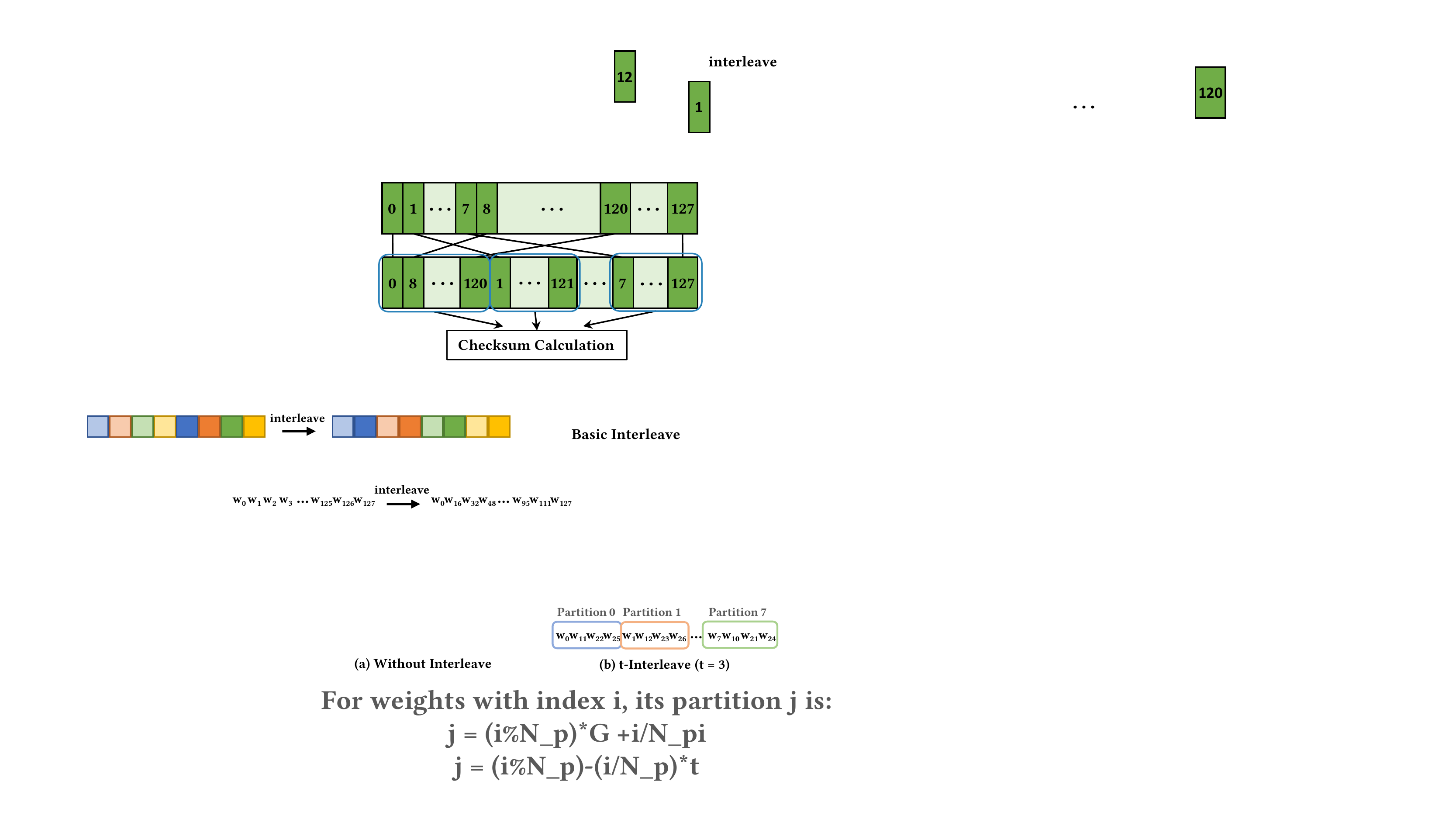}
  \caption{Basic interleaving strategy in checksum calculation. The checksum is calculated on a group of interleaved weights.}
  \label{fig:fig4-5}
\end{figure}

The basic interleaving process is shown in \cref{fig:fig4-5}. For the case when there are 
$N$ groups where each group consists of weights that were originally $N_W$ locations apart, the $k$-th group consists of weights $k + N_W\times l$, where $0 \leq l <N$ and $0 \leq k< N_W$. So for $N= 16$, $N_W=8$, group $0$ consists of weights in locations $0,8,16, \ldots, 120$ as shown in the figure. We choose $N_W=G$ and add an additional offset of $3$ in all our experiments. 


The interleaving distance can be  kept as secret and stored in the secure on-chip SRAM. It can be different from one layer to the next making it even harder for the attacker to know which two bits are in the same group. We will show that the interleaving strategy  not only addresses the security concern, but also improves the detection of multiple bit-flips. 

\begin{algorithm}[t] 
\caption{Pseudo-code of signature calculation}
\label{alg:loop}
\begin{algorithmic}[1]
\Require{8-bit fixed point weight tensor $B_{i}$ in layer $i$. $B_{i,j}$ denotes $j$th group of weights  in layer $i$,  $K_i$ is the secret key for layer $i$, and $N_W$ is the parameter for  interleaving. $N$ denotes the total number of groups of size $G$ in layer $i$.}

\Ensure{Signature $S_{i,j}$}
\Statex
\Function{SignCal}{$B_{i},~G$}
        \State $B_{i}^*$ $\gets$ Interleave ($B_{i}$, $N_W$) \Comment{Interleaving}
        
        \For{~$j$ = 0~:~$N$}
            \State$B_{i,j}^*[t]$ $\gets$ $  B_{i}^*[j\times G: (j+1)\times G]$ \Comment{Grouping}
            \For{~$t$ = 0~:~$G$}
                \State sign $\gets$ $K_i.next()$ \Comment{Secret Key}
                \If {~sign $==$ 0~}:
                \State$B_{i,j}^*[t]$ $\gets$ $ - B_{i,j}^*[t]$     \Comment{Two's Complement}
                \EndIf
            \EndFor
          \State $M$ $\gets$ $\sum_{t=0}^{G-1} B_{i,j}^*[t]$\Comment{Summation}
          \State $S_{i,j}$ = Binarize(M, 2) \Comment{Signature}
          
      \EndFor
\State \Return {$S_{i}$}
\EndFunction
\end{algorithmic}

\end{algorithm}
\subsection{Overall Algorithm}

The overall detection scheme is described in Algorithm 1. The weights in a layer, $B_i$, are reorganized into groups with weights that are originally interspersed. 
The secret key is applied on the interleaved weights to determine the sign of each weight (referred to as masking) in the checksum computation. The 2-bit signature for each group is the binarized summation of the scaled weights.
The signatures $S_{i,j}$ of each group $B_{i,j}$ in $B_i$ are calculated and stored as golden signature in on-chip memory. During run-time, a fresh set of signatures is computed for every new chunk of data that is fetched from the cache. The detection is performed by comparing the computed signature with the golden signature.

\section{RADAR Scheme: Recovery}
If an attack does occur, a successful detection can help halt the system to stop making decisions, wait for downloading a clean copy of weights or let the secondary system take over. This may result in significant increase in timing overhead so next we describe a very simple recovery scheme that can recover most of the model accuracy instantly.

In the PBFA analysis described in Section III.C, we see that PBFA attacks the MSB of a small weight and converts  it to a large value. It is this large change in weight value that causes a direct change in the ReLU activation and thus has a dramatic effect on the output. So we locate the groups where the bit-flips have occurred using the proposed fine-grain detection scheme and then set all the weights in that group to 0. A de-interleaving step is required when  interleaving is applied prior to checksum calculation during the weight update so that the original weight organization is not affected. 
Since most of the weights in a group have small values and are clustered around 0, setting all the weights in the group to 0 works well especially if the partition size is small. In Section VI we demonstrate that this scheme helps regain most of the accuracy lost due to PBFA for ResNet-18 and ResNet-20 models.





\section{Experiments}
\subsection{Settings}
We demonstrate the effectiveness of our detection and protection scheme for image classification using two popular datasets:  CIFAR-10~\cite{krizhevsky2014cifar} and ImageNet~\cite{deng2009imagenet}. 
We use a 8-bit quantized ResNet-20 for CIFAR-10 dataset and 8-bit quantized ResNet-18 for ImageNet~\cite{he2016deep}.
ResNet-20 model is trained from scratch for 200 epochs using Adam, with a learning rate of 0.01 and decay of 0.0001. We use a pre-trained model for ResNet-18 and fine-tune it for 20 epochs using SGD.




\subsection{Detection Performance}




\vspace{-5pt}
%

\begin{figure}[ht]
  \centering
  \includegraphics[width=1.0\linewidth]{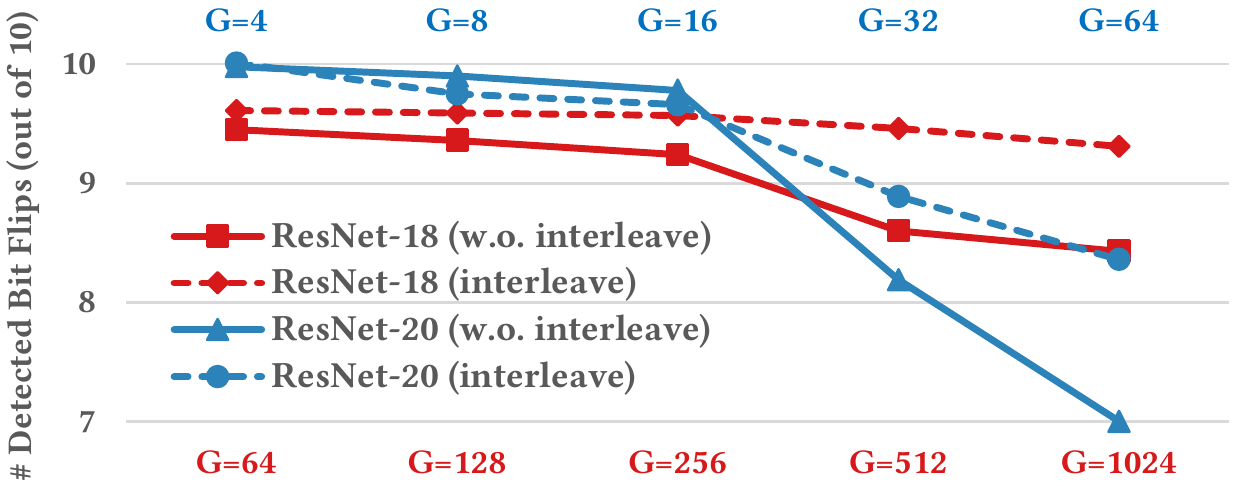}
  \caption{Average number of detected bit-flips (out of 10 bit flip attacks) using PBFA.
  The group size G is swept from 4 to 64 for ResNet-20 model and 64 to 1024 for ResNet-18.} 
  \label{fig:figB}
  \vspace{-5pt}
\end{figure}

We set the number of bit-flips to be equal to 10 per attack since this is sufficient to cause a significant performance degradation and calculate the number of bit-flips that were detected. We perform each  attack 100 times. The detection performance with and without the interleaving strategy is shown in Fig.~\ref{fig:figB}. For the ResNet-20 model (shown in blue), the detection performance without interleave approaches 10/10 when $G$ is small, and  drops to around 7/10 when $G=64$. This is consistent with the observation in Fig.~\ref{fig:figA} where we show that when $G$ is large, the proportion of multiple bits in the same group sharply increases thereby harming the detection performance. With interleaving, the detection performance for large group size is better because of the reduction in the number of multiple bit-flip cases. For the ResNet-18 model (shown in red), we observe that interleaving results in a very high 9.5/10 detection ratio even  when the group size is large.

We also investigate the probability of failing to detect an attack on the MSBs. We consider a layer with 512 weights and
perform $10^6$ rounds of bit-flips with 10 random bit-flips on MSB position per round. We find that for group size $G=32$, the detection miss rate is $10^{-5}$; for group size $G=16$, the detection miss rate further reduces to $10^{-6}$. Since the number of weights in a group is much larger than the toy example, we conclude that the proposed detection scheme exposes even smaller risk for full-fledged networks.

\subsection{Recovery Performance}
To evaluate recovery performance, we consider number of bit-flips (N\_BF) of 5 and 10.  For ResNet-20 and ResNet-18 we compare choices of different group sizes with and without interleaving. Recall that in the proposed recovery technique, the  weights in a group are set to zero if a bit-flip is successfully detected. So in this experiment, we check the test accuracy of the revised model obtained by setting the weights of a group to 0 if that group detected bit-flips.      

\begin{table}[htbp!]
  \caption{Accuracy Recovery of the RADAR scheme}
  \label{tab:accu}
  \setlength\extrarowheight{1pt}
  \resizebox{0.96\columnwidth}{!}{%
  \begin{tabular}{|c|c|c|c|c|c|}
  \hline
   \multirow{2}{*}{Model}& \multirow{2}{*}{$N_{BF}$} & \multicolumn{4}{c|}{Test Accuracy (\%)}\\
  \cline{3-6}
  &  & Baseline & \multicolumn{3}{c|}{w.o. interleave/ with interleave} \\
  \cline{1-6}
  \multirow{3}{*}{ResNet-20}& 0& 90.15 &G = 8 & G = 16 & G = 32 \\
     & 5& 40.72 &82.66/85.64 & 76.39/83.72 & 68.06/73.35\\
     & 10 & 18.01 & \textbf{80.86/81.07} & 70.53/77.96 &61.62/61.32\\
    \hline
    \multirow{3}{*}{ResNet-18}&0& 69.79&G = 128 & G = 256 & G = 512\\
    & 5& 5.66 &66.60/67.51 & 65.12/66.15 & 62.89/62.87\\
     & 10 & 0.18 & 62.69/66.33 & 59.95/64.96 &\textbf{57.46/60.69}\\
    \hline
  \end{tabular}
  }
\end{table}

\begin{figure}[ht]
\vspace{-5pt}
  \centering
  \includegraphics[width=1.0\linewidth]{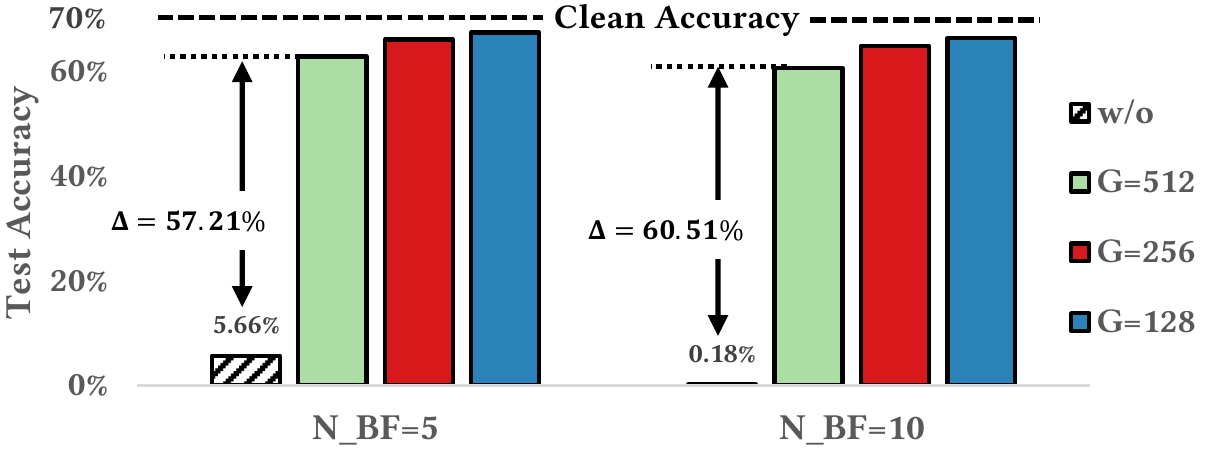}
  \caption{Accuracy recovery performance on ResNet-18 model (ImageNet).}
  \label{fig:FigE}
  \vspace{-5pt}
\end{figure}

The results for ResNet-20 and ResNet-18 are shown in Table~\ref{tab:accu}. For ResNet-20, the accuracy after the attack drops to as low as 18.01\% with N\_BF=10. After recovery, the accuracy can climb up to 81\% when $G=8$ and 62\% when $G=32$. Similarly for ResNet-18, the accuracy drops to 0.18\% with N\_BF=10 and climbs to 66\% when $G=128$ and 61\% when $G=512$.
We see that the accuracy recovery is consistently better when the interleaving strategy is used. Fig.~\ref{fig:FigE} further illustrates the test accuracies for ResNet-18. While there is a mild performance drop when $G$ increases, the accuracy recovery is consistently quite high.


\section{Discussion}







\subsection{Tradeoff between recovery \& storage}
We use gem5 \cite{binkert2011gem5} to evaluate the timing overhead of RADAR. We use a 8-core processor build in gem5. Each core is instantiated as an Arm Cortex-M4F core and runs at 1GHz frequency. The system is equipped with a two-level cache: L1-32KB and  L2-64KB.
The layer information and weights are obtained from the pre-trained ResNet-20 and ResNet-18 models. Our detection and recovery procedure, RADAR, is embedded in the computations of every layer. For RADAR scheme with group size $G$, we use padding if the number of weights in a layer is not divisible by $G$. 

\begin{figure}[ht]
\vspace{-5pt}
  \centering
  \includegraphics[width=1.0\linewidth]{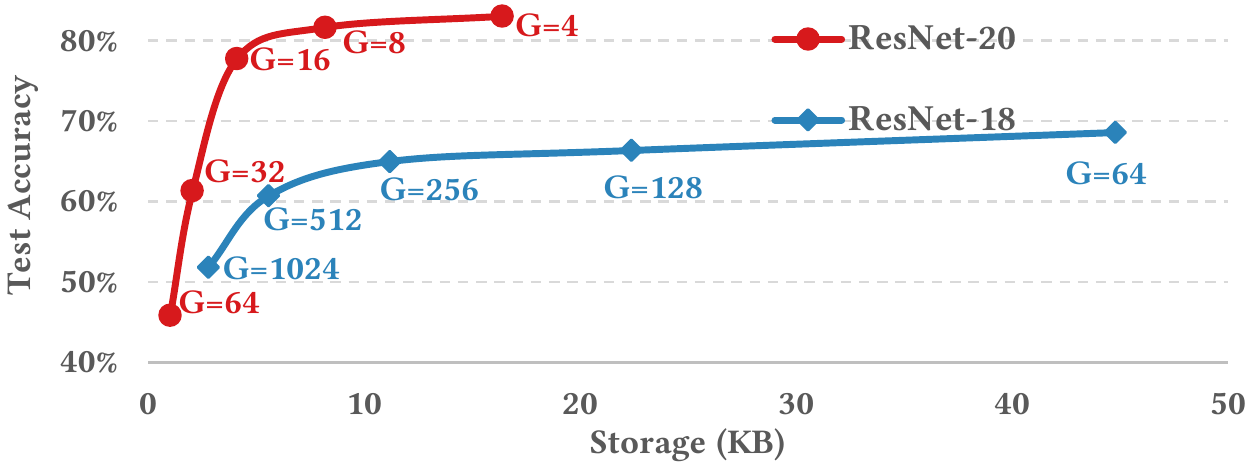}
  \caption{Test accuracy after recovery  vs. Storage overhead of proposed RADAR scheme under PBFA with N\_BF = 10 on ResNet-20 and ResNet-18 models.}
  \label{fig:fig8}
  \vspace{-5pt}
\end{figure}

To choose a good group size, we study the tradeoffs between recovery performance and storage overhead. Fig.~\ref{fig:fig8} plots recovery accuracy as a function of storage overhead for ResNet-18 and ResNet-20 models. For ResNet-20, the best accuracy-storage tradeoff occurs at $G=8$. The accuracy under 10 bit-flips is still over 80\% and the storage overhead for the signature bits is 8.2 KB, which can be easily stored on-chip. For ResNet-18, $G=512$ seems to be the best choice. The accuracy can be kept at over 60\% for 10 bit-flip attack and the storage overhead is just 5.6 KB. 


\begin{table}[ht]
\caption{Time Overhead of RADAR}
\vspace{-5pt}
\setlength\extrarowheight{1pt}
\centering
\begin{tabular}{ |c|c|c|c| } 
 \hline
 {} & Original & RADAR & Overhead  \\ 
 \hline
 ResNet-20& 66.3ms & 68.7ms (69.8ms) & 3.56\% (5.27\%)  \\ 
 ResNet-18& 3.268s & 3.287s (3.328s) & 0.58\% (1.83\%) \\ 
 \hline
\end{tabular}
\label{tab:time}
\vspace{-5pt}
\end{table}

The time overhead for RADAR on ResNet-20 with $G=8$ and ResNet-18 with $G=512$ for batch size of 1 is shown in Table~\ref{tab:time}. The time overhead with interleaving is shown in brackets. The overhead is quite small -- 5.27\% for ResNet-20 and 1.83\% for ResNet-18  with interleaving. The time overhead can be further reduced in a multi-batch inference setting, where each chunk of weights is loaded once and used many times.








\subsection{Comparison with Related work}


RADAR has been designed to address the strongest adversarial weight attack to date, i.e., PBFA, via a fast and light weight checksum algorithm that has a very low storage overhead.
Other options to perform single and double bit-flip detection for general data integrity checking include Cyclic Redundancy Check (CRC) \cite{koopman2004cyclic} and Hamming Code \cite{hamming1950error} based Double-bit Error Detection. To provide for recovery, both codes require significantly higher storage overhead. For instance Hamming code requires 7 bits for 64 bits of data (corresponding to group size of 8) and 13 bits for 4096 bits (corresponding to group size of 512). Similarly, to achieve a HD=3, CRC needs 7 bits and 13 bits for group size of 8 and 512 respectively. 


We compare the performance of RADAR with the competitive CRC schemes. Table~\ref{tab:compare} shows the total inference time, the overhead time for detection ($\Delta$) and the storage overhead for ResNet-20 when G=8 and  ResNet-18 when G=512.
For  ResNet-18, when G=512, CRC-13 has a time overhead if 0.317s compared to 0.060s. The storage overhead of CRC-13 is 36.4KB, compared to 5.6KB which is required by RADAR.  If only the MSBs were to be protected, we would require CRC-10 which has a time overhead of 0.315s and storage overhead of 28.0KB, which is still significantly larger than RADAR.

\begin{table}[ht]
\vspace{-5pt}
\caption{Overhead comparison with CRC techniques}
\vspace{-5pt}
\setlength\extrarowheight{1pt}
\resizebox{0.96\columnwidth}{!}{%
\begin{tabular}{|c|c|c|c|c|} 
 \hline
 \multirow{2}{*}{Scheme} & \multicolumn{2}{c|}{ResNet-20; G=8} & \multicolumn{2}{c|}{ResNet-18; G=512}  \\ 
  \cline{2-5}
  & Time/$\Delta$ &Storage & Time/$\Delta$ &Storage\\
 \hline
 CRC & 84.2ms/17.9ms & 28.7KB & 3.585s/0.317s & 36.4KB\\ 
 \hline
 RADAR & 69.8ms/3.5ms & 8.2KB & 3.328s/0.060s & 5.6KB\\
  \hline
\end{tabular}
}
\label{tab:compare}
\vspace{-10pt}
\end{table}

\section{Knowledgeable Attacker}
Next we assume that the attacker knows that a checksum-based scheme has been used to detect attacks on MSBs.
However, the attacker does not know the secret key that is used for masking the weights  and/or the interleaving strategy. 

\begin{figure}[ht]
  \centering
  \includegraphics[width=0.95\linewidth]{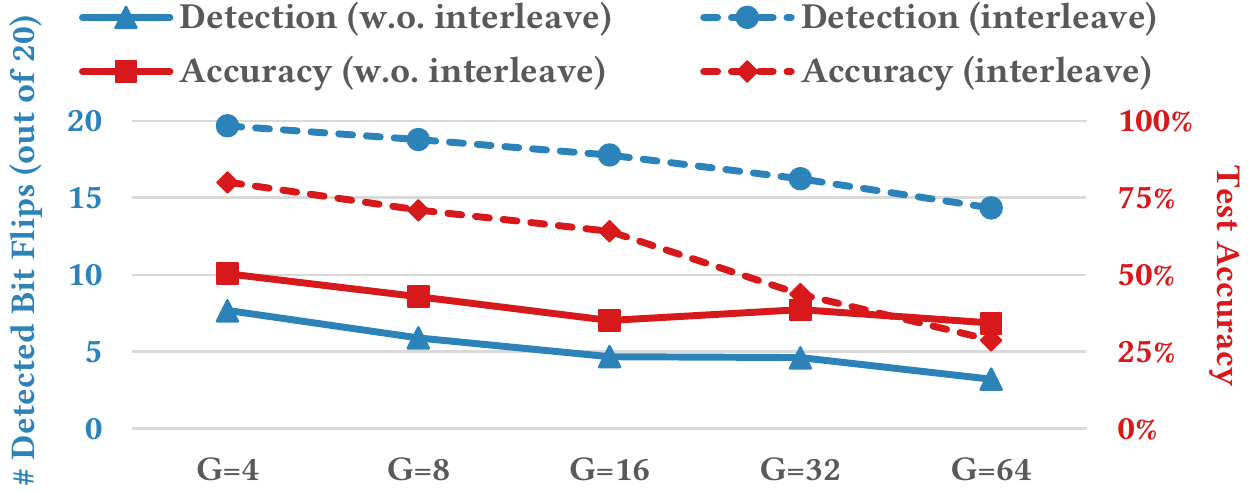}
  \caption{Detection and accuracy recovery performance on ResNet-20 model against knowledgeable attackers.}
  \label{fig:figC}
  \vspace{-5pt}
\end{figure}

\textbf{Flip multiple bits in a group.} In addition to attacking the 10 bits identified by PBFA, the attacker could add in another 10 bits to flip (20 bit-flips in total). These bit flips would be of the form (0 $\rightarrow$ 1 and 1 $\rightarrow$ 0) to evade detection. As shown in Fig.~\ref{fig:figC}, the detection performance without interleaving drops greatly (lower blue line) causing the accuracy recovery to be low as well. By applying interleaving, the detection ratio can be keep at a level similar to the traditional PBFA case. Also, the accuracy is much higher when group size is small.

\textbf{Avoid flipping MSB.} 
The proposed checksum-based detection scheme with a 2-bit signature cannot detect bit-flips on less significant bits as effectively as it can on MSB. 
However,  many more bits are required to launch a successful attack if only MSB-1 or lower significant bits are allowed to be flipped. For instance, the attacker needs around 30 bit-flips  on MSB-1 bits (compared to 10 bit-flips on MSB) for comparable accuracy degradation on the ResNet-20 model. We address attacks on MSB-1 bits by utilizing a 3-bit signature computed by  binarizing $M$ to 3 bits.
While this method has a higher storage overhead (3-bit signature vs 2-bit signature), it can successfully detect errors due to attacks on MSB-1 bits.

\section{Conclusion}
In this work, we propose RADAR, a low overhead run-time adversarial weight attack detection and accuracy recovery scheme for PBFA. We show that the RADAR scheme has very low timing overhead and can be embedded in the  inference process.
A thorough analysis of the PBFA attack characteristics helped derive a simple error detection and accuracy recovery scheme. A 2-bit signature is computed using addition checksum of a group of interspersed weights and compared with the golden signature to
determine whether a PBFA attack had been launched on that group or not. 
We show that RADAR has superior detection performance; it can consistently detect over 9.5 bit-flips out of 10 injected bit-flips.
A simple but effective accuracy recovery scheme is built on top of this detection scheme. It can recover the accuracy from 0.18\% back to above 60\% for a ResNet-18 model on ImageNet. The gem5 simulation shows that for ResNet-18 model, the RADAR scheme only increases the inference time by $<$2\%, making this scheme highly suitable for run time detection and protection.



\bibliographystyle{unsrt}
\bibliography{reference_short}

\end{document}

%% file: math_commands.tex
\usepackage{amsmath,amsfonts,bm}

\def\eqref#1{equation~\ref{#1}}

\def\1{\bm{1}}

\DeclareMathAlphabet{\mathsfit}{\encodingdefault}{\sfdefault}{m}{sl}
\SetMathAlphabet{\mathsfit}{bold}{\encodingdefault}{\sfdefault}{bx}{n}

